\documentclass[showpacs, prl,twocolumn,fixfloat,a4paper,superscriptaddress]{revtex4}
\usepackage{epsfig,bm}
\begin{document}

%\hyphenation{Bran-den-bur-gi-sche}
\title{The strength of the radial-breathing mode in single-walled carbon nanotubes}

\author{M. Mach\'on}
\affiliation{Institut f\"ur Festk\"orperphysik, Technische Universit\"at Berlin, Hardenbergstr. 36, 10623 Berlin, Germany}

\author{S. Reich}
\affiliation{University of Cambridge, Department of Engineering, Trumpington Street, Cambridge CB2 1PZ, UK}

\author{H. Telg}
\affiliation{Institut f\"ur Festk\"orperphysik, Technische Universit\"at  Berlin, Hardenbergstr. 36, 10623 Berlin, Germany}

\author{J. Maultzsch}
\affiliation{Institut f\"ur Festk\"orperphysik, Technische Universit\"at Berlin, Hardenbergstr. 36, 10623 Berlin, Germany}

\author{P. Ordej\'on}
\affiliation{Institut de Ci\`encia de Materials de Barcelona, Campus de la U.A.B., 08193 Bellaterra, Barcelona, Spain}

\author{C. Thomsen}
\affiliation{Institut f\"ur Festk\"orperphysik, Technische Universit\"at Berlin, Hardenbergstr. 36, 10623 Berlin, Germany}

\begin{abstract}
We show by \emph{ab initio} calculations 
that the electron-phonon coupling matrix element $\mathcal{M}_{e\mbox{\emph{-}}ph}$ of the radial breathing mode in single-walled carbon 
nanotubes depends strongly on tube chirality. 
For nanotubes of the same diameter the coupling strength $|\mathcal{M}_{e\mbox{\emph{-}}ph}|^2$  
is up to one order of magnitude stronger for zig-zag than for armchair tubes. For ($n_1$,$n_2$) tubes $\mathcal{M}_{e\mbox{\emph{-}}ph}$ depends on the value of $(n_1-n_2)\,\mathrm{mod\,} 3$, which allows to discriminate semiconducting nanotubes with similar diameter by their Raman scattering intensity.
We show measured resonance Raman profiles of the radial
breathing mode which support our theoretical predictions.
\end{abstract}


\maketitle
The radial breathing mode (RBM)  is without doubt the best known feature in the Raman spectra of
carbon nanotubes. In this vibration all carbon atoms move in the radial direction
creating a breathing-like deformation of the entire tube. This mode is unique to single-walled carbon
nanotubes and is taken as indicative of the presence of nanotubes in a sample.
Moreover the frequency of the radial breathing mode is proportional to the inverse diameter of the 
tube~\cite{rao97}. Raman scattering is therefore often used to determine the diameter 
or diameter distribution in nanotube samples~\cite{bandow98,milnera00,jorio01}. 
In detail, the relation between nanotube diameters in real samples and the radial 
breathing mode spectrum is more complicated, because of the resonances in the Raman process 
and additional force constants coming from the tube-tube van-der-Waals interaction in bundled 
nanotubes~\cite{milnera00,venkateswaran99,thomsen99a}. Furthermore, the RBM eigenvector has a small non-radial component~\cite{dobardzic03,kuerti03}.

It was suggested to use the RBM to find not only the tube diameter but also 
the chiral angle, \emph{i.e.}, to identify a particular $(n_1,n_2)$ nanotube
~\cite{jorio01,kramberger03,bacsa02}. $n_1$ and $n_2$ specify the chiral vector 
$\bm{c}=n_1\bm{a}_1+n_2\bm{a}_2$ around the circumference of a nanotube in terms of the 
graphene unit cell vectors $\bm{a}_1$ and $\bm{a}_2$. This assignment relied mostly on the 
frequency of the RBM, sometimes combined with an argument about the resonant 
enhancement of the Raman intensity for the laser excitation energy~\cite{jorio01}. It was, however, always assumed that 
the electron-phonon coupling of the RBM is independent of the chirality of a 
tube~\cite{richter97}. This means that far from resonance or exactly in resonance the scattering intensity of the 
radial breathing mode is expected to be the same for different $(n_1,n_2)$ nanotubes.
Only recently, Strano \emph{et al.}~\cite{strano03} 
suspected a smaller matrix element for armchair tubes than for zig-zag tubes 
from their measurements of the RBM signal strength of a series of 
carbon nanotubes in solution.

In this article we show that, 
contrary to the wide-spread assumption,
the electron-phonon coupling strength of the radial breathing mode 
depends on the diameter \emph{and} chirality  of the nanotube. 
In \emph{ab initio} calculations we 
find the  squared electron-phonon
matrix elements in zig-zag tubes to be up to one order of magnitude stronger than in armchair tubes
for the same optical transition energy. In 
semiconducting nanotubes the matrix elements 
allow to distinguish between the 
$(n_1-n_2)\,\mathrm{mod\,} 3=\pm1$ nanotube families. A similar intensity
difference is expected for the two transitions of metallic nanotubes in each branch of the
Kataura plot~\cite{kataura99}. We show experimental evidence of this intensity difference based
on measurements of resonant Raman profiles of the RBM of nanotubes in aqueous solution. 
The relative Raman intensities can  independently  confirm an
$(n_1,n_2)$ assignment obtained, \emph{e.g.},
by photoluminescence. 

In the expression for the Raman-scattering cross-section from perturbation
theory the square of the electron-phonon matrix elements
$|\mathcal{M}_{e\mbox{\emph{-}}ph}|^2$ appears in the numerator. The intensity
of the Raman signal in full resonance is scaled by the electron-phonon coupling~\cite{trallero92}.
When calculating these matrix elements both electrons and holes must be taken into 
account. To every electron excited into a conduction band $c$ and interacting with a 
phonon, corresponds a hole in the valence band $v$. Adding up the two
contributions, {\it i.e.}, assuming strict electron-hole symmetry, we obtain
for the electron-phonon matrix  element
\begin{equation}
\mathcal{M}_{e\mbox{\emph{-}}ph}=\langle{\bf k}c|H_{e\mbox{\emph{-}}ph}|{\bf k}c\rangle-\langle{\bf k}v|H_{e\mbox{\emph{-}}ph}|{\bf k}v\rangle,
\label{cond_val}\end{equation}
where  the minus sign comes 
from the opposite charges of holes and electrons. 

The diagonal matrix elements of the electron-phonon coupling Hamiltonian $H_{e\mbox{\emph{-}}ph}$
for optical phonons 
can be obtained from the shift of the electronic bands under deformation of the atomic structure 
corresponding to the phonon-pattern \cite{khan84}
\begin{equation}
\langle{\bf k},b|H^i_{e\mbox{\emph{-}}ph}|{\bf k},b\rangle=\sqrt{\frac{\hbar}{2MN\omega_{\mathrm{RBM}}^i}}
\sum_a\epsilon^i_a\frac{\partial E_b({\bf k})}{\partial {\bf u}_a},
\label{khan}
\end{equation}
where the sum runs over all atoms in the unit cell. {\bf k} and $b$ denote, 
respectively, the wave vector and band index of the electronic state, $i$ indexes the phonon,  
$M$ is the atomic mass, $\epsilon^i_a$ 
the polarization vector of the phonon, normalized as
$\sum_a\epsilon^i_a\epsilon^j_a=\delta_{ij}$ , $E_b({\bf k})$ the electronic
energy and {\bf u}$_a$ the atomic displacement. $N$ is the number of unit
cells in the system (1 in our calculation).

We calculated the $\Gamma$-point phonon spectrum and the band structure of several 
isolated  nanotubes in their minimum-energy configuration and under deformation due 
to the RBM to obtain the change of the electronic energies  in 
Eq.~(\ref{khan})~\cite{verissimo01}.
The RBM eigenvector was obtained from finite
differences; it has a small
non-radial component~\cite{dobardzic03,kuerti03} which lowers the calculated
matrix elements by $\approx$~30\,\%.    
%%%%%%%%%%%%%%%%%%%METHOD%%%%%%%%%%%%%%%%%%%%%%%%%%%%%%%%%%%%%%%%%%%%%%%%%
All calculations 
were performed with the SIESTA code~\cite{soler02} within the local
density approximation~\cite{perdew81}. The core electrons were replaced by non-local
norm-conserving pseudopotentials \cite{troullier91}. A double-$\zeta$,
singly polarized basis set of localized atomic orbitals was used for the valence electrons, with
cutoff radii of 5.12\,a.u. for the $s$ and 6.25\,a.u. for the $p$ and $d$ orbitals~\cite{junquera01}. 
16 $k$ points in the $k_z$ direction were included for metallic nanotubes and 3 $k$
points for semiconducting tubes. Real-space integrations were done in a grid with a cutoff
$\approx$ 270~Ry. 
%%%%%%%%%%%%%%%%%%%%%%%%%%%%%%%%%%%%%%%%%%%%%%%%%%%%%%%%%%%%%%%%%%%%%%%%%%%%%%%%%%%%%%%%%%%%%%%%%%%%

\begin{figure}
\epsfig{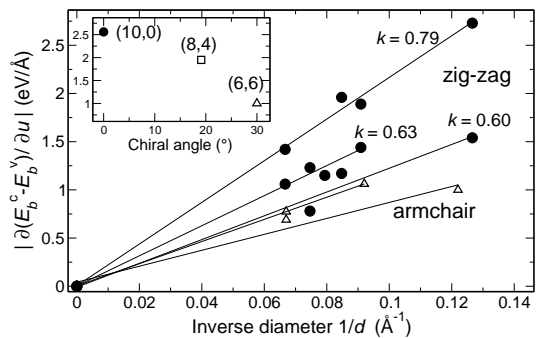}
\caption{\label{comp} Calculated band-energy changes (absolute value) per unit change in radius
for zig-zag (circles) and armchair tubes (triangles). Lines are linear
fits of the data corresponding to the indicated $k$ value (as defined in the
horizontal axis of Fig.~\ref{graph}c); they include the origin as a data point.
{\bf Inset}: Band-energy change $|\partial E_b({\bf k})/\partial {\bf u}|$ for the first optical transition of nanotubes with diameter $\approx 8$~\AA{} as a function of chiral angle. }
\end{figure}
In Fig.~\ref{comp} we show the energy change 
 $\partial E_b({\bf k})/\partial {\bf u}=\partial [E_b^c({\bf k})-E_b^v({\bf k})]/\partial {\bf u}$ for several tubes as a function of  inverse tube diameter.
The data was evaluated 
at the region of the Brillouin  zone with the highest optical transition probability, \emph{i.e.}, the band extrema~\cite{reich02}.
We grouped the data corresponding to $k$ points that are close when mapped onto the graphene Brillouin zone (lines in Fig.~\ref{comp}). To first approximation, the points of a particular group correspond to the same transition energies. We find that for a particular transition energy, $|\partial E_b({\bf k})/\partial {\bf u}|$ is proportional to $1/d$ and tends to zero in the infinite-diameter limit.
This trend can be easily understood, since the  same change in radius yields smaller
bond-length changes for bigger tube radii. The infinite-diameter limit
corresponds to a translation of graphene,
which cannot affect the electronic system.
It is clearly seen that $|\partial E_b({\bf k})/\partial {\bf u}|$ for zig-zag tubes is up to 2.5 times
larger than for  armchair tubes.
The inset of Fig.~\ref{comp} shows the energy change for the first transition of tubes all with diameter  $\approx$8~\AA{} but different chiralities; in particular, the (8,4) tube lies  between the armchair and the zig-zag values. 
The assumption of a
chirality-independent electron-phonon interaction  is thus
incorrect. Our results suggest the use of relative Raman
\emph{intensities} for discriminating chiralities.

\begin{ruledtabular}
\begin{table*}
\begin{tabular}{lcccccccccccc}
     &  (6,0)  &   (10,0) &  (6,6)&  (8,4)& (11,0)&(8,8)&   (14,0)&   (15,0)&   (16,0)   &   (17,0)&   (11,11)&  (19,0)     \\
\hline
$V_c\,($\AA$^3)$&77.2   &210  &130    &626&  254      &230    &  406&    467&    534  &       603&    436&    755     \\
d\,(\AA)& 4.8  &    7.9  &    8.2 &   8.4  &  8.7     &10.9 &    11.0 &    11.8 &    12.6    &   13.4  &     15.0 &  15.0        \\
$\omega_{ph}$\,(cm$^{-1}$)
      &   446  &    287  &   278  &     274& 257&    209  &     203 &    188  &     179    &   170   &      151 &  149         \\
\hline
$\mathcal{M}_1$   
    &{\bf 0.050}&{\bf-0.028}& -0.015 & -0.013&{\bf 0.021}&-0.010 &{\bf 0.016}&{\bf-0.022}&{\bf-0.017}&{\bf 0.014}&-0.005&  {\bf-0.015}  \\
&       (1.0)   &(0.8)  &(2.3)  &(0.8) &{\bf (0.9)} &(1.8)  &       (0.7)   &(1.5)  &(0.6)  &       (0.6)   &(1.4)  &(0.5)  \\
\hline
$\mathcal{M}_2$   
   &{\bf-0.062}&{\bf 0.017}&  --  &  0.004&{\bf -0.028}&-0.015  &{\bf-0.020}&{\bf 0.013}&{\bf 0.013}&{\bf-0.016}&-0.010&{\bf 0.012}  \\
&       (1.7)   &(2.0)  &       &(1.7)&{\bf (1.3)}  &(3.1)  &       (1.1)   &(2.0)  &(1.2)  &       (1.0)   &(2.5)  &(1.0)  \\
\hline
$\mathcal{M}_3$   
   & {\bf --} &{\bf-0.030}&   --  &  -0.016& {\bf -0.028}&-0.017  &{\bf-0.021}&{\bf-0.022}&{\bf-0.018}&{\bf-0.017}&-0.012& {\bf-0.016}  \\
&               &(2.4)  &       &(2.6)&{\bf (2.6)}&(3.7)  &       (2.4)   &(2.6)  &(1.9)  &       (2.1)   &(3.3)  &(1.6)  \\
\hline
$\mathcal{M}_4$   
   & {\bf --} &{\bf-0.031}&   --  &   --& {\bf -0.028}&   --    &{\bf --   }&{\bf-0.022}&{\bf  -- } &{\bf 0.009}&   --  & {\bf-0.016}\\
&               &(3.0)  &       &       &  {\bf (3.1)}&     &               &(3.2)  &                       &(2.4)&         &(2.7)  \\      
\end{tabular}
\caption{\label{table.rbm.suma}  Calculated diameters, RBM frequencies, and
electron-phonon matrix elements $\mathcal{M}_{e\mbox{\emph{-}}ph}$ (in eV) 
for the first optical transitions (\emph{ab initio} calculated energies in eV in parenthesis~\cite{reich02}). The matrix elements (bold for zig-zag tubes) were rounded to 0.001 eV. Rows labeled $\mathcal{M}_{1\mbox{-}4}$ correspond to the first four optical transitions
for light polarized parallel to the nanotube axis. A `--'  indicates a lack of the transition or
a band shift which could not be evaluated for technical reasons. 
Note that $\mathcal{M}_{e\mbox{\emph{-}}ph}$ is normalized to the unit-cell volume $V_c$.
}
\end{table*}
\end{ruledtabular}
In Table~\ref{table.rbm.suma}, the calculated matrix elements and RBM frequencies are summarized; we found  $\omega_{\mathrm{RBM}}\approx C_1/d+C_2$ 
with $C_1=232$~cm$^{-1}$nm  in
 excellent agreement with the literature~\cite{kuerti98,sanchez99a} and $C_2=-6~$cm$^{-1}$.
The largest difference in the matrix elements $\mathcal{M}_{e\mbox{\emph{-}}ph}$ between
zig-zag (bold face) and armchair nanotubes of similar diameter and, hence,
same RBM frequency is found for the (11,11) and the (19,0) tubes. The matrix
element $\mathcal{M}_3$ of the (19,0) tube is by a factor of three larger than
$\mathcal{M}_1$ of the (11,11) tube although the two transition energies are similar (see Table~\ref{table.rbm.suma}). Since the Raman signal is proportional to $|\mathcal{M}_{e\mbox{\emph{-}}ph}|^2$, we expect the RBM intensity to be nine times larger for the (19,0) than for the (11,11) nanotube from the difference in $|\mathcal{M}_{e\mbox{\emph{-}}ph}|$ alone. For different transition energies this ratio could be even larger.  

\begin{figure}
\epsfig{file=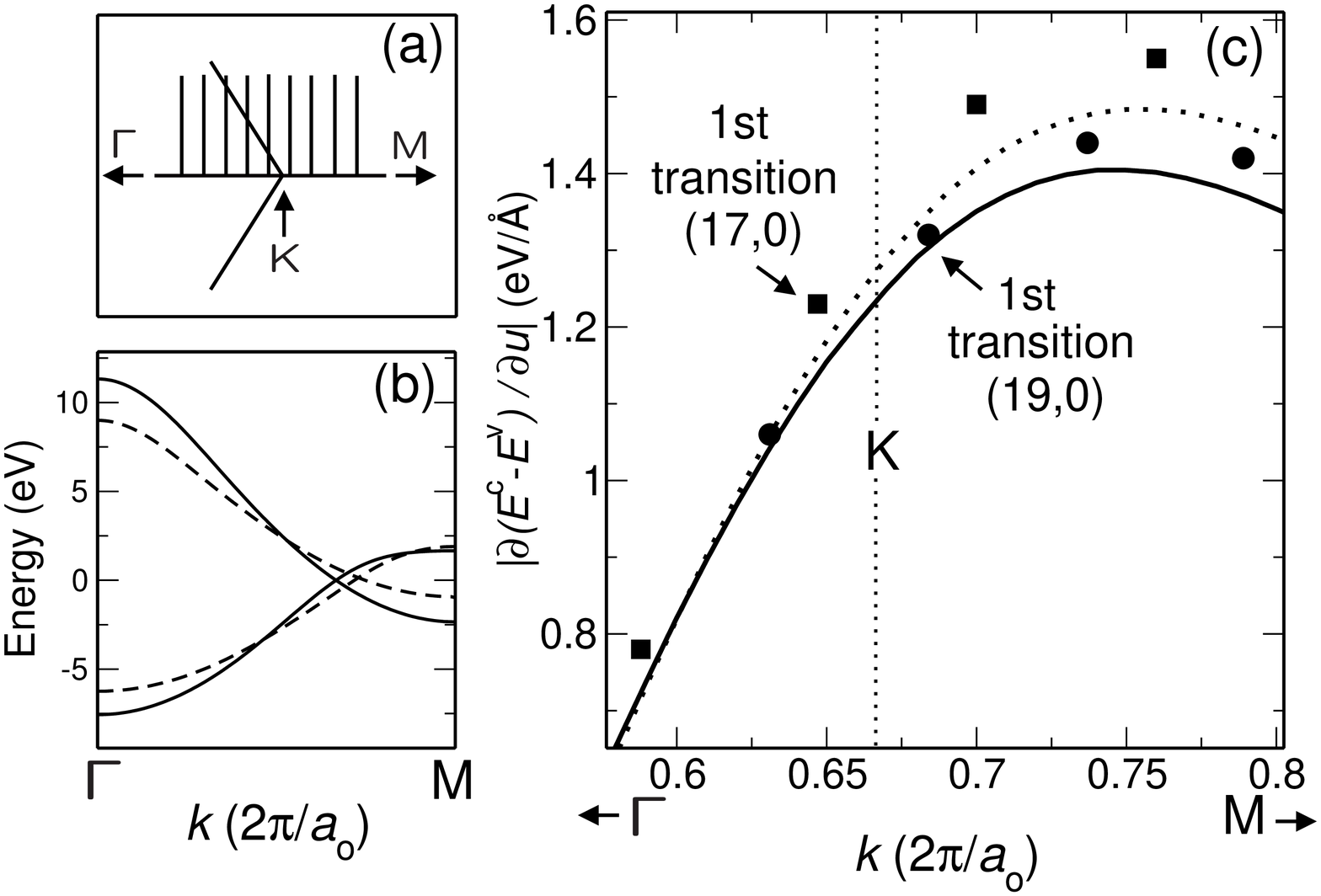,width=7cm,clip=}
\caption{\label{graph} a) Detail of the Brillouin zone (BZ) of graphene around $K$ with the band lines of a (19,0) nanotube and the $\Gamma$-$K$-$M$ line. b) Electronic bands of graphene along $\Gamma$-$K$-$M$ in equilibrium (solid line) and under a deformation of 0.1~\AA{} corresponding to the RBM of a (19,0) tube (dashed, difference enhanced $\times  10$). c) Calculated $\partial E_b({\bf k})/\partial {\bf u}$ for a
(17,0) (squares) and a (19,0) (circles) tube, and
of graphene deformed to simulate the RBM of a (19,0)
tube (solid line) and a (17,0) nanotube (dotted). 
}
\end{figure}

The matrix elements of zig-zag tubes show another interesting feature:
they have either a larger magnitude  and are negative or
a smaller magnitude and are positive [\emph{e.g.}, for the (10,0) tube
$\mathcal{M}_1$=$-$0.028 and $\mathcal{M}_2$=$+$0.017]. A change in sign is very uncommon in electron-phonon interaction
in solid-state systems. The matrix elements are positive in most
semiconductors~\cite{cardona82}. To explain this unusual behavior
 we calculated $\partial E_b^{c,v}({\bf k})/\partial {\bf u}$ 
of a graphene sheet ($a_0^{\mathrm{equil.}}=2.467$~\AA{}) stretching the sheet in the zig-zag direction to simulate the radial atomic displacement and adding the non-radial component (see Fig~\ref{graph}). 

In Fig.~\ref{graph}c we show $|\partial E_b({\bf k})/\partial {\bf u}|$ for graphene when stretching it according to a (19,0) tube
(solid line) and a (17,0) tube (dotted) together with the \emph{ab initio}
calculated values for these two tubes. 
$\mathcal{M}_{e\mbox{\emph{-}}ph}$ depends in sign and magnitude on where the
 optical transition occurs with respect to the $K$-point of graphene.
The $\Gamma$-point states of an $(n,0)$ zig-zag tube in the graphene BZ 
are obtained by dividing the $\Gamma$-$K$-$M$ line into $n$ parts (see
Fig.~\ref{graph}a). The states closest to the $K$-point of graphene have the lowest  transition energies. The (17,0) tube, {\it e.g.}, has its 1$^{st}$ transition to the 
left of the $K$-point  and the 2$^{nd}$ one to the right. The energy shift of the graphene bands is smaller to the left of the K point than to its right (Fig.~\ref{graph}c). Therefore for this tube $|\mathcal{M}_1/\mathcal{M}_2|<1$. The (19,0) tube, on the other hand, has its 1$^{st}$ transition to the right of the $K$-point and the 2$^{nd}$ to the left, 
yielding $|\mathcal{M}_1/\mathcal{M}_2|>1$. Furthermore, $\partial E_b({\bf k})/\partial {\bf u}$ is negative to the right  of the $K$-point, and positive to the left, explaining the signs of $\mathcal{M}_{e\mbox{\emph{-}}ph}$. 
In general, all semiconducting tubes can be divided into  $\nu=(n_1-n_2)\,\mathrm{mod\,} 3=\pm1$ families, which  behave like the (17,0) and (19,0) tube with respect to sign and relative magnitude of the $\mathcal{M}_{e\mbox{\emph{-}}ph}$.
Metallic nanotubes  usually have two close-by transition energies due to 
trigonal warping~\cite{reich00}. The transition with lower  energy
originates from the right of the $K$-point, the one with higher energy  from its left. Therefore,
the lower-energy  transition is expected to give a higher Raman intensity.

%%%%%%%%%%%%%%%%%%%%%%%%%%%%%%%%%%%%%%%%%%%%%%%%%%%%%%%%%%%%%%%%%%%%%%%%%%%%%%%%%%%%%%%%%%%%%%%%%%
%%%%%%EXP%%%%%%%%%%%%%%%%%%%%%%%%%%%%%%%%%%%%%%%%%%%%%%%

To confirm our theoretical predictions we performed Raman scattering measurements on
nanotubes in solution~\cite{bachilo02,telg04}. Raman spectra were excited with a
Ti-Sapphire laser, recorded with a DILOR XY800 spectrometer, and
corrected for the sensitivity of the experimental setup. We then calculated
the squared scattering amplitudes $|W_{FI}|^2$ from the  Raman signal.

\begin{figure}
%\begin{center}
\epsfig{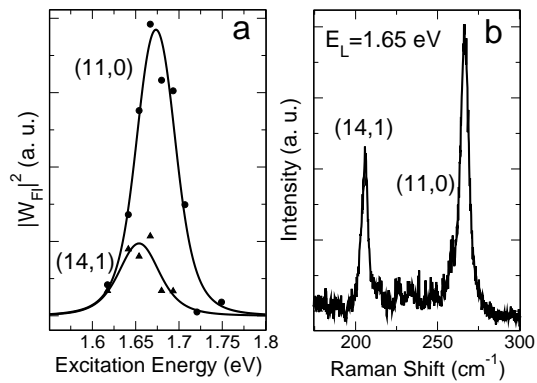}
%\resizebox{0.68\columnwidth}{!}{
%\includegraphics*{raman_abs_inset2.eps}}
%\end{center}
\caption {{\bf a)} Measured resonance Raman profiles (symbols) and fits (lines)
of two different nanotubes. {\bf b)} Raman spectrum with a laser energy of 1.65\,eV.
 } \label{daten}
\end{figure}

In Fig.~\ref{daten}a we show two selected resonance profiles of radial
breathing modes. Using the assignment by Bachilo \emph{et
  al.}~\cite{bachilo02} we
identify these resonances as the second transition of the (14,1) 
nanotube with $\nu=+1$ and the (11,0) tube with $\nu=-1$~\cite{fussnote}. We predicted a
higher Raman intensity for nanotubes with $\nu=-1$ in excellent agreement
with the experimental data. As shown in
Fig.~\ref{comp} $\partial E_b({\bf k})/\partial {\bf u}$ is similar for nanotubes with similar diameter,
chiral angle, and $\nu$. Approximating the matrix element for the (14,1)
tube by $\partial E_b({\bf k})/\partial {\bf u}$ for the (16,0) nanotube we find theoretically  $|\mathcal{M}_{2}^{(11,0)}/\mathcal{M}_{2}^{(14,1)}|^2\approx$~3.
Experimentally, the ratio 
$|W^{(11,0)}_{FI}/W^{(14,1)}_{FI}|^2\approx$~4 is  in excellent agreement with our \emph{ab-initio} result and a uniform distribution of chiral angles in nanotube samples~\cite{henrard99}. We
stress that the intensity difference between $\nu=-1$ and $+1$ tubes is
generally observed in our experiment and not limited to the two profiles
shown. A more detailed study is
underway, but beyond the scope of this paper.

Raman-based $(n_1,n_2)$ assignments performed so far relied strongly on a Raman
intensity analysis~\cite{jorio01} that implicitly assumed constant electron-phonon
coupling. As shown, this is not correct. 
The most intense peak does not necessarily correspond to the nanotube
closest to resonance. 
As an example,  we show in Fig.~\ref{daten}b a Raman spectrum for an excitation energy of
1.65~eV. This excitation energy corresponds to the maximum in the resonance
profile of the (14,1) nanotube, whereas the (11,0) nanotube is excited 25~meV
below its resonance. The Raman intensity thus depends not only on the resonance
condition, but also on the particular tubes under study. Differences in
Raman intensity due to resonances cannot be distinguished from the chiral
angle dependence of $\mathcal{M}_{e\mbox{\emph{-}}ph}$ using only a single excitation energy; instead a
resonance profile has to be evaluated.

A chirality-dependent electron-phonon coupling naturally explains the observations
by Strano~\emph{et al.}~\cite{strano03}. They mapped the electronic transitions
of metallic nanotubes using Raman excitation profiles.
Strangely, the
armchair tubes were apparently missing in their  sample. 
In contrast, photoluminescence experiments on the same type
of sample~\cite{bachilo02} suggested a predominance of
large-chiral-angle tubes. Our calculations solve this apparent contradiction: the Raman signal of armchair nanotubes is small due to a weak
electron-phonon coupling. 

In conclusion, we calculated the electron-phonon matrix elements for carbon nanotubes. The matrix elements of zig-zag tubes are much larger than those of  armchair tubes, leading to
a larger  Raman signal for smaller chiral angle tubes. 
Furthermore, for semiconducting tubes the magnitude and the sign of 
the matrix elements change systematically for different
optical transitions and $\nu=\pm 1$ nanotubes.
Relative Raman
intensities can discriminate between armchair and zig-zag tubes as well as 
$\nu=\pm 1$ tube families. The latter we demonstrated by measuring
the radial breathing mode resonance on a $-1$ and $+1$ nanotube. The family
and chiral-angle dependence of the Raman intensities can be used for a
refined assignment of chiral indices and chirality distributions.

We thank F. Hennrich for providing the samples. S. R. acknowledges financial support  by the Berlin-Bran\-den\-bur\-gi\-sche Akademie der
Wissenschaften, the Oppenheimer Fund, and Newnham College. J. M. was supported by the DFG (Th 662/8-2).
We acknowledge the MCyT (Spain) and the DAAD (Germany) for a Spanish-German Research action; P. O. acknowledges support from Spain's MCyT grant BFM2003-03372-C03-01.

\end{document}